\documentclass [12pt] {article}
\usepackage{amssymb,amsfonts,amsmath}
\usepackage{url}
\usepackage{graphicx}
\usepackage{comment}
\usepackage[utf8]{inputenc}
\usepackage[russian,english]{babel}
\usepackage{geometry}
\newcommand{\cnd}{\mskip 1mu|\mskip 1mu}

\let\le=\leqslant
\let\ge=\geqslant
\newcommand{\rus}[1]{\foreignlanguage{russian}{#1}}

\usepackage{color}

\def\K{{\rm K}}

\title{Kolmogorov complexity in the USSR (1975--1982): isolation and its end\thanks{Translated from Russian by Alexander Shen.}}

\author{V.V.V'yugin\thanks{
Institute of information transmission problems, Russian Academy of Sciences.}}
                    
\date{}
\begin{document}
\maketitle

\begin{abstract}
These reminiscences are about the ``dark ages'' of algorithmic information theory 
in the USSR. After a great interest in this topic in 1960s and the beginning of 1970s 
the number of people working in this area in the USSR decreased significantly. At that time 
L.A.~Levin published a bunch of papers that were seminal for the modern algorithmic 
information theory. Then he left the USSR, and the new wave of interest was triggered 
by the talk of A.N.~Kolmogorov at a Moscow  State (Lomonosov) University Mathematical Department 
(Logic and Algorithms Division) seminar organized by him; several younger 
researchers obtained some new results in algorithmic information theory. 
\end{abstract}

\section{Leonid Levin} 

I met Leonid Levin during the academic year 1971/2. At that time I was a first-year 
graduate student and lived (as well as many of them) in a dormitory called 
``The Graduate Students and Interns House'' (Dom aspiranta i stazhera). 
This was a large dormitory; each room was occupied by four or five persons
The inhabitants were a mixture of graduate students 
from all the departments of the Moscow State University, interns, visitors and 
many other strangers.

Leonid (Lenya) Levin graduated from the Moscow State University mathematics department 
(``mekh-mat'') a year before myself, in 1970. He was not accepted to the graduate 
school despite the support of Andrei Kolmogorov and despite of his obvious talents 
and achievements (e.g., he already published a survey~\cite{ZvL70} about Kolmogorov 
complexity, with many new results, together with Alexander Zvonkin). This was 
impossible due to the veto of the Russian secret service, KGB, and the communist party 
commitee (because of political reasons; moreover, as if it were not enough, Levin was of 
Jewish origin). He managed to get some temporary positions due to the help of 
Kolmogorov (if I remember correctly, Kolmogorov managed to obtain an intern position 
in his laboratory for Levin) and others (e.g., the vice-director of the Institute of 
Information Transmission Problems, Iosif Abramovich Ovseevich, secured later a temporary 
position in the Institute for Levin). From time to time one could see Lenya waiting in 
the queue of visitors to the dormitory manager when he tried to extend his official stay 
in the dormitory. On the other hand, one could illegally live there with some friends 
(if one has enough of them).

Lenya was acquainted, it seems, with all the graduate students of the math department (and some students 
from other departments) who lived in this dormitory, including myself. Everybody knew him, and we all 
called him Lenya. His had extremely wide interests at the time. It was difficult to find some topic 
in science (and not only in science) that was of no interest to him, and he had a strong opinion on most topics. He discussed virtually all mathematical results obtained at the time.

Moreover, often people from different applied fields came to Lenya and discussed their work. 
This (volunteer) consulting sometimes provided solutions for their problems.

At that time we never discussed Kolmogorov complexity. But I remember that Lenya told me about 
his (strong) results connected with step-counting functions, 
Rabin's compression theorem and Blum's speedup theorem. These were a hot 
topic at the time. These results were published together with the Russian translation of one of 
Blum's papers~\cite{Lev74a}. He told me about these results.

The topic of my graduate work (1971--1974) was the semilattice of computable numbering of 
recursively enumerable sets. (Now these sets are called `computably enumerable'.) The notion 
of a computable numbering was introduced by Kolmogorov, and his student of 50s, Vladimir Uspensky, 
was my thesis advisor. In 1970s this topic was intensively developed in Novosibirsk Institute of 
Mathematics. People there were interested in my work; it became a topic of several expository 
talks at their ``Algebra and logic'' seminar. 
   
I did not learn about Kolmogorov complexity and algorithmic randomness before 1975. However, 
I remember being (in 1970 or 1971) in a rather crowded room; it was a seminar organized by Kolmogorov, 
and more than 30 people were there (except for Kolmogorov, I remember also Levin, Max Kanovich and 
Nikolai Petri). Probably this was a predecessor of the Kolmogorov complexity seminar organized 
by him later in the beginning of 1980s.

I remember Kolmogorov's attitude at the time: he considered Levin as a colleague rather than 
a student or a young researcher. I remember an occasion when Kolmogorov and Levin discussed some 
problem that Levin was working on; this problem was just solved by somebody else. Kolmogorov said 
to Levin something like ``we are mature enough mathematicians not to be disappointed when somebody 
else was the first to solve the problem''.

One more reminiscence: during the seminar Kolmogorov asked Levin whether he could give a talk about 
something. Levin answered (in his style) that yes, he is able to give a talk on any topic at any 
time and in any place.

My contacts with Levin continued in 1975. At that year I finished my thesis,  defended it and was 
sent for obligatory work to Bashkir State University in Ufa (a center of Bashkiria region in the USSR; 
after graduation, Soviet students have to work for three years at the place chosen by the authorities, 
only later they were allowed to change the job). There were good mathematicians in Ufa, but they were 
mostly interested in mathematical analysis or differential equations, and I had no chances for 
scientific collaboration there.

I visited Moscow quite often, and once met with Levin at the Moscow State University. 
He invited me to the apartment he rented in Moscow. It was near the subway station called ``Kakhovskaya'', 
near the street called ``Chernomorsky bulvar'' (boulevard).

At that time he got some better position in some applied research institute where he has enough free 
time to work on his own topics. He had some Soviet-reasonable salary and was able to rent an apartment 
in Moscow. At the same time his life was quite stressful. He had (and not without a reason!) a permanent 
feeling that at any moment the Soviet secret services could interfere and force him to be fired and sent 
out of Moscow.

Lenya liked visitors and was quite happy if they stayed with him for a while.  I think that many people 
(of approximately the same age as him) remember the apartment he rented, his hospitality, nice and 
joyful people and interesting talks (often oppositional to the authorities, as it was then taken among 
the intelligentsia). Often foreign visitors joined the company. I remember 
a funny philosophical discussion of freedom and non-freedom in the USSR. The American guest, 
a woman named Judy, who visited the USSR several times and liked the joyful companies and discussions, 
noted that freedom is an internal feeling. Lenya replied with a (metaphorical?) story. Peter Gacs 
(his Hungarian friend and colleague since 1970s, now they work both in Boston University) comes 
to Moscow with very tasty hungarian sausage roll. Each day he cuts and eats a small piece of the roll. 
When the sausage is over, he goes back to Hungary (that was considered at the time as a bit more 
developed country than USSR).

During the summer of 1975 I came to Moscow for a long time, and during July we made long walks with Lenya. 
While walking in the Bitsa park (a rather large park that was on the boundary of Moscow at that time), 
we discussed almost everything: politics, philosophy, mathematics, universal prediction and universal 
semimeasure, the notion of information content for infinite sequences. Sometimes Albert Abramovich Muchnik 
(of Muchnik -- Friedberg fame) joined our discussions.

Levin was a fast thinker and usually was ahead in the conversation and could convince other participants. 
However, it was easy to speak with him. He never tried to enforce his opinion upon the others. 
He just provided rational arguments that usually convinced others without pressing them to 
accept Levin's viewpoint. This made him a popular person and many people had high regard for him. 
It seems that local party bosses and the KGB understood this and were suspicious about possible 
influence of Levin. People like him had no place in the USSR.

In 1975 Levin formulated a question that he considered as philosophically important. 
After some simplification, this problem was stated as follows. Consider infinite binary 
sequences modulo Turing reductions, i.e., Turing degrees. One can say that two mutually 
reducible sequences can be transformed one into another by a computable operators 
(oracle machines), and therefore ``contain the same information'' 
(up to the finite amount of information hidden in the reductions). We consider 
Borel sets (properties) of infinite sequences that are invariant under these reductions, 
i.e., unions of arbitrary families of Turing degrees. One of these invariant families 
is formed by (Turing degrees) of Martin-L\"of random sequences. Levin showed earlier 
that one can consider only sequences that are random with respect to the uniform 
Bernoulli measure: every sequence that is Martin-L\"of random with respect to some 
computable measure, has the same degree as some sequence that is Martin-L\"of random with 
respect to the uniform measure~\cite{ZvL70}. The only exceptions are computable sequences. 
They are random with respect to computable measures where they are atoms
(have positive measure), but are not Turing-equivalent to sequences that are Martin-L\"of random 
with respect to the uniform measure (those sequences cannot be computable for obvious reasons). 
The computable sequences form another invariant class. Now the question is whether they exist 
other non-trivial invariant classes.

Of course, this question should be formulated more precisely to be interesting. 
Note that the union of an arbitrary family of Turing degrees is an invariant class (if measureble). 
For example, one may consider just one non-computable Turing degree, i.e., the set of all sequences that are 
Turing-equivalent to a given non-computable sequence. However, such a property is 
not ``real'' in a sense that no probabilistic process (algorithm with access to a
fair coin) producing a finite or infinite binary sequences can produce a sequence 
with this property (from a fixed non-computable Turing degree) with positive probability. 
Indeed, the famous theorem of de Leuuw, Moore, Shannon and Shapiro~\cite{LMSS} guarantees 
this.  There are other invariant sets of sequences with the same property of 
being ``non-real'' (no probabilistic process can generate their elements with positive 
probability). Levin gave a description for classes with this property. Namely, for each 
set with 
this property there is some sequence $\alpha$ such that every sequence from the set contains 
infinite amount of information about $\alpha$. Moreover, for each $\alpha$ the set of 
sequences 
that contain infinite amount of information about $\alpha$, has this property 
(see~\cite{Lev74}).\footnote{Levin gave some definitions of mutual information in two 
infinite sequences in this paper and subsequent ones. 
The quantity of information $I(\alpha:\beta)$ should have an invariance property: 
it cannot be increased by random or deterministic (recursive) processing of $\alpha$ 
or $\beta$ (see, e.g.,~\cite{Lev84}). He insisted that this 
notion is very important and different approaches to its definition should be developed.}
In this way we get invariant sets of sequences whose universal measure (maximal 
lower semicomputable continuous semimeasure, called also a priori probability) is zero.\footnote{
The semicomputable universal (a priori) measure was introduced in~\cite{ZvL70}. The a priori measure of 
a Borel set of infinite sequences is equal to the probability that the corresponding probabilistic 
Turing machine will produce a sequence from this set.} 
Levin called sets with this property \emph{negligible}. 

We want to ignore negligible sets. Formally, consider the equivalence relation on 
invariant sets: two sets are equivalent if their symmetric difference is negligible. 
In this way we get a structure that was called \emph{the algebra of invariant properties}. 
The operations in this algebra are restrictions of the usual set-theoretic 
operations to invariant sets. The minimal element is the equivalence class $\mathbf{0}$ of 
negligible sets. The maximal element is the class $\mathbf{1}$ of all sequences. There are two 
naturally defined elements in this algebra: the class $\mathbf{r}$ generated by Martin-L\"of random 
sequences, and the class $\mathbf{c}$ generated by computable sequences.

Levin told me that the elements $\mathbf{c}$ and $\mathbf{r}$ are atoms of the algebra of invariant 
properties, i.e., cannot be represented as a union of two other non-zero elements from this algebra.

Levin~\cite[definition 4.4]{ZvL70} called ``proper'' (this term is used in the English translation of 
that article; Russian name was \rus{<<правильная последовательность>>}) or ``complete''~\cite{LeV77,Lev84} 
the sequences that are Martin-L\"of random with respect to some computable measure. 
Sequences that are Turing-equivalent to complete sequences are called regular (see~\cite {Lev84}).

Every regular sequence $\alpha$ is either computable or Turing-equivalent to some sequence $\beta$ that 
is random with respect to uniform Bernoulli measure (see~\cite{ZvL70}). The prefixes of $\beta$ have 
monotone complexity 
equal to their length (up to a constant). Therefore one could say that information contained 
in a regular sequence can be compressed maximally. The property of being atoms 
(for $\mathbf{c}$ and $\mathbf{r}$) can be interpreted as follows: the only invariant information 
about a regular sequence is the amount of information in it that can be either finite ($\mathbf{c}$) 
or infinite ($\mathbf{r}$). The information in the elements of $\mathbf{r}$, after being compressed 
optimally, is indistinguishable from random noise.

Levin then asked me: is it true that there are no other elements (except for the elements generated 
by computable and Martin-L\"of random sequences, and, of course, $\mathbf{0}$ and $\mathbf{1}$) in 
this algebra, i.e., $\mathbf{c}\cup \mathbf{r}=\mathbf{1}$, or some other elements exist?

When I returned to Ufa from Moscow (in the summer of 1975),  I managed to construct a counterexample 
to the conjecture that no other elements exist, i.e., a probabilistic algorithm whose output is with 
positive probability (that can be made close to $1$ but not equal to $1$) a noncomputable sequence 
that is not Turing-equivalent to any Martin-L\"of random sequence. 
Therefore, $\mathbf{c}\cup \mathbf{r} \ne \mathbf{1}$ and the algebra contains more elements 
(not only the four elements mentioned), see~\cite{Vyu76}.

From a philosophical viewpoint one could say that there is a natural process that with positive 
probability generates arrays of information that cannot be ``explained'' as 
being random with respect to some computable measure. One can even say that these non-regular sequences 
cannot be optimally compressed without infinite information loss. This can be considered as an application 
of computability theory to the foundations of probability theory (and foundations of mathematics in general).

As we already mentioned, Levin has proven that $\mathbf{c}$ and $\mathbf{r}$ are atoms in the algebra of 
invariant properties (cannot be presented as non-trivial unions of smaller elements). 
Levin then asked: what about other atoms in this algebra? It is easy to see that there is at most 
countable set of atoms. Approximately in 1977--1978 I managed to construct indeed a countable family 
of atoms, and an infinitely divisible element (that has no atoms as subsets). 
Therefore, we have the disjoint union
$$
\mathbf{c}\cup \mathbf{r}\cup\bigcup_{i=1}^\infty \mathbf{a}_i\cup \mathbf{d}=\mathbf{1},
$$
where $\mathbf{a}_i$ are atoms that do not contain computable sequences, and $\mathbf{d}$ is 
the rest that is the infinitely divisible element. It seems that this result has no impact 
on further research, and it is a pity that the properties of these $\mathbf{a}_i$ are still not studied.

At that time Levin was developing a general theory of random objects. He wanted to define uniform 
randomness deficiency (with respect to semimeasures, i.e., measures on the set of finite and infinite 
sequences) in such a way that every infinite binary sequence would be random with respect to the maximal 
semimeasure (now also sometimes called ``the continuous a priori 
probability'').\footnote{Some results in this direction were later published in~\cite{Lev84}.} 
He also dreamed about a universal language that would allow us to give concise definition 
for mathematical objects and functions. The hope was that by using this language one could make 
the hidden constants in the statements about Kolmogorov complexity reasonable. 
We discussed these questions again and again in 1975--1978.

Levin also told me about the ``brute force search problem'' (<<problema perebora>> 
in Russian) that later become famous as $\text{P}=\text{NP}$ question. He stated 
the problem in a short note~\cite{Lev73} published in ``Information Transmission 
Problems'' journal; this note contained several examples of NP-complete problems. 
It went almost unnoticed at that time, but later became rather famous. The history of the early attempts 
to approach this problem in the USSR is explained in an excellent survey by Boris 
Trakhtenbrot~\cite{Tra84}. In the same paper Levin noted that there 
is an algorithm (sometimes called ``L-search'') that is optimal in the sense that 
if there is an efficient (e.g., polynomial) algorithm for NP-complete problems, 
then Levin's algorithm is almost as fast. (Of course, the catch is that nobody knows 
whether this algorithm is polynomial.) The argument that Levin had in mind used some 
new version of Kolmogorov complexity that takes into account not only the program size 
but also the (logarithm) of its computational time. However, the paper was rather short 
(as usual for Levin), and the result was stated not only without proof, but also without 
the description of the search algorithm. That time Levin told me about this algorithm and the corresponding
complexity.\footnote{Later this algorithm was presented in~\cite{Lev84}, see also the survey~\cite{Vyu99}.}

Levin suggested that I should work on $\text{P}=\text{NP}$ question: at that time the situation does 
not look as hopeless as now. Probably it was a good luck for me that my personal life has distracted 
me from concentrating on this problem.

In 1978 Levin decided to emigrate 
to the US using the Israel invitation as a tool (as many people did: getting an invitation to repatriate 
to Israel was an unreliable but essentially the only way to get out of the USSR). It is remarkable 
that Levin waited so long. 
This way out of the USSR was possible (at least for some people, mostly Jews) already for several years, 
and the KGB people who summoned Levin gave advices to leave Moscow, e.g., to go to Armenia (that was 
at that time a part of the USSR). This could be a hint that Levin emigration would be desirable 
for KGB and he would be allowed to leave.

At that time Soviet border, like Styx, was essentially one-way. The emigrant should renounce 
the Soviet citizenship (i.e., write a letter asking for that; during some periods, there was 
a quite significant fee for such a procedure). Of course, there were other problems, 
like language barrier. We both knew enough English to read a mathematical paper almost without 
a dictionary, but Levin's knowledge of Russian language and culture was magnificent: for example, 
he knew a lot about Alexander Blok, Russian poet (``Silver Age'' of Russian culture, the beginning 
of XX century) and could read by heart his poems for hours.

During our cooperation in 1975--1978 Lenya was quite isolated in a scientific sense; he cannot 
go to conferences and seminars abroad (the authorities definitely would 
not let him out, so he did not even try). His PhD thesis (under the supervision of 
Kolmogorov) was failed in Novosibirsk 
despite of the positive reports of all the reviewer and the advisor (Kolmogorov). 
(See more details in~\cite{BS2009}. The translated version of this thesis was published in~\cite{Lev2010}.)

Once Levin was invited to a computer science conference (Mathematical Foundations of Computer Science)
in Czechslovakia; at  that time it was a Soviet-controlled country (so authorities could not worry 
about a chance to escape) but nevertheless none of us could go 
there officially. (The invitation was sent to Levin by some member of the organizing committee. 
He suggested that I could try to go there, but my Ufa bosses told me that I have no chance --- one 
should get 18 different kinds of permissions for that.) Still we sent a paper~\cite{LeV77} there, 
though it was forbidden at the time (one has to apply for a special permission to send the paper abroad; 
for that a special official certificate was needed that confirms that ``the paper does not contain 
anything that could be considered as a discovery or an invention''). This well could end in firing 
from the job (or even worse), but the Soviet system was quite old at the time and couldn't keep track 
of all violations of that (relatively small) scale.

Of course, Levin has many friends with whom he could discuss scientific questions, 
including already mentioned Alexander Zvonkin (the coauthor of the survey~\cite{ZvL70}), Albert Muchnik, 
Peter G\'acs. And, of course, Levin was always supported by Kolmogorov. As an academician, 
Kolmogorov was able to endorse Levin's publications in the \emph{Reports of the Academy of Sciences} 
journal (``Doklady AN SSSR''), and quite a few Levin's papers were published there. 
Levin was also able to publish papers in another reputable journal, \emph{Information Transmission Problems}, 
published by the IITP (Institute for Information Transmission Problems).

However, this was more a hobby than a regular activity in the framework of some academic or research 
institution. The problems of the algorithmic information theory were not discussed at the university,
in particular, Levin could not be a student supervisor. The algorithmic information theory was 
``a marginal'' science.  

It is not a coincidence that Levin published more during a difficult periods of his life. 
Here are some papers that cover his main results in the algorithmic information 
theory:~\cite{Lev73a,Lev74,Lev76,Lev76a,Lev76b,Lev76c,Lev77}.

It seems I was the only collaborator of Levin from the USSR working with him on algorithmic information 
theory. But Peter Gacs that came several times from Hungary worked with Leonid and obtained significant 
results in the field. 

Looking back, I feel that I was lucky that Levin has so much time to discuss mathematics 
(and other topics) with me. Were he a prominent and recognized scientist in the USSR (at the level of his 
international recognition later), he would be rather busy and we would have much shorter and much more 
formal meetings.

Trying to follow Judy's example and find some advantages of Soviet life, I could say that the management 
of our official jobs did not care much about what we are doing and often allowed us to do what we wanted 
not caring about the results, there was no pressure to get papers published, to write grant applications 
and reports, etc. Having few career opportunities, we could be ``mentally free people'' and could work 
for a long time on some abstract (almost philosophical) problems --- or just relax and do nothing.

In 1977 I got married and was allowed to come to Moscow. Looking for a job in Moscow was a hard task. 
It was impossible to get a teaching position in some higher education institution: there were many 
qualified people who did not want to lose their right to live in Moscow by going elsewhere. 
The only jobs available were the so-called ``otraslevye instituty'', centers organized by government to work on 
the problems of specific industry.  I changed three jobs of this type. In the first and the second 
ones I was able to work only for a year, or, better to say, to be paid. The ``work'' there was 
completely disorganized, people did not do anything (and often knew nothing about the industry 
they were assumed to help), the management gave idiotic orders, etc. On the other hand, people 
with PhD, including myself, were given a good (according to Soviet standards) salary. 
At that time I tried to ignore the surrounding farce and finish my paper about invariant properties.

In 1978 Lenya and his wife Larissa got a permission to emigrate and went to Vienna by a train 
(from the ``Belorusskaya'' railroad station). Many people came to this station to say goodbye 
to them. When leaving, Lenya told me that he believes that algorithmic information theory will 
get more attention in the USSR. He also wrote a reference and gave a specific advice: 
to show my paper about the algebra of invariant properties to Kolmogorov, and gave me his phone number.

\begin{center}
\includegraphics[width=0.9\textwidth]{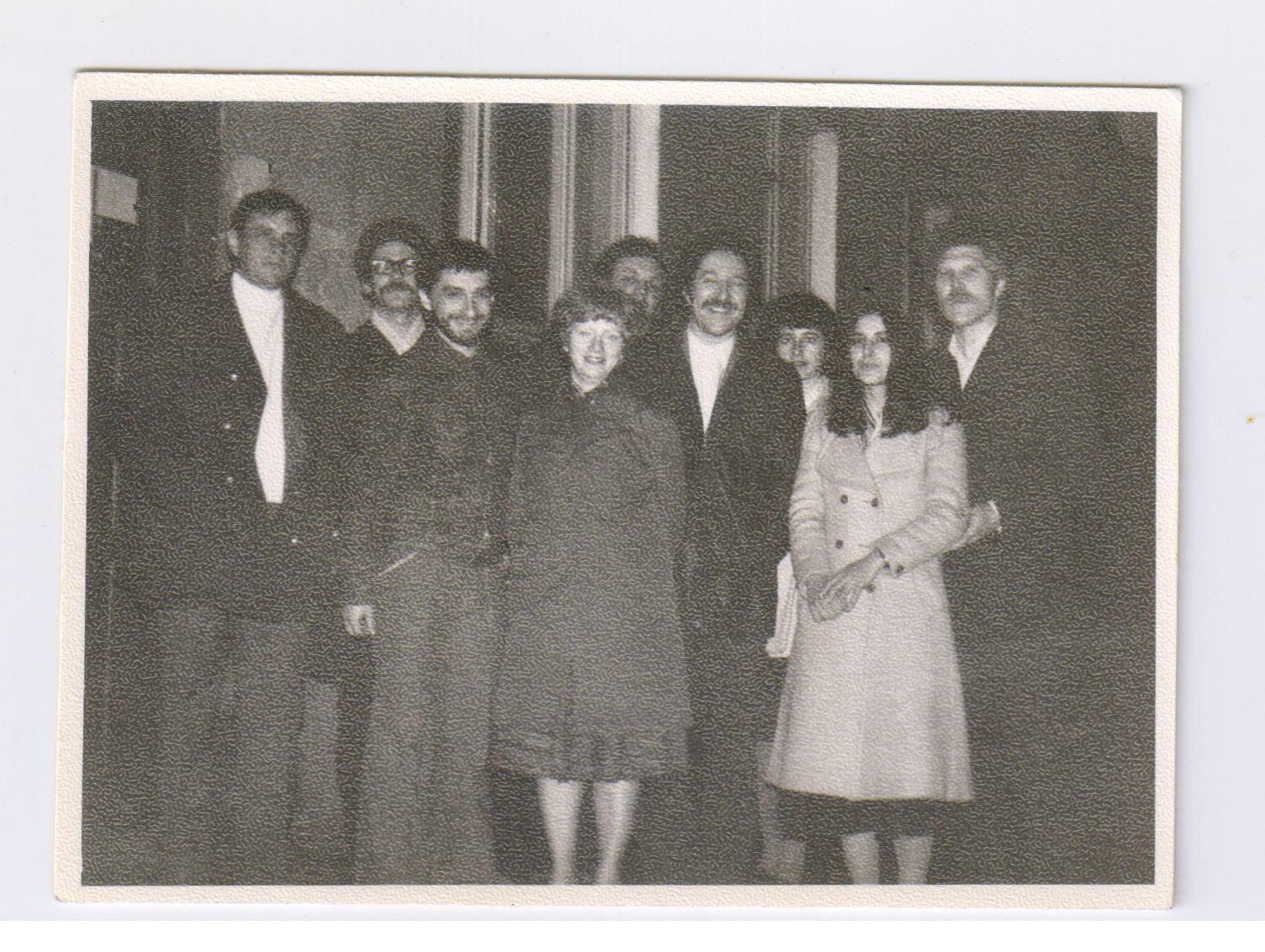}\\
Leonid Levin (3rd from the left) and his wife Larissa (8th from the left)
and their friends just before their departure from the USSR (also Eduard~Dumanis
(5) and his wife (4), Vladimir V'yugin (5) and his wife Elena (7), 
Albert Muchnik (9))
\end{center}

\section{Andrei Nikolaevich Kolmogorov}

When my paper was ready, approximately in 1979, I phoned Kolmogorov and said that Lenya Levin suggested 
that I show my paper on the algorithmic information theory to him (Kolmogorov). Kolmogorov gave me his 
address (he lived in the building of the Moscow State University, near the department of mathematics) 
and told me when to come. When I arrived, he opened the door, asked who I am and took my paper. 
He asked me to call him again in two weeks. When I did that, he started to apologize that he has no time 
to read the paper, and asked me to call again in two weeks. The same thing repeated for a long time, 
it took may be a year and a half. Of course, sometimes I waited longer than two weeks, sometimes 
Kolmogorov was ill for few months, etc.

Levin was right predicting that interest in Kolmogorov complexity would revive. 
My thesis advisor, Vladimir Uspensky, asked me (around 1980) to write a survey about 
Kolmogorov complexity and algorithmic randomness. I never tried to approach systematically 
the notions of algorithmic information theory, but I remembered our discussions with Levin 
(while walking), also his papers of 1973--1977 (mostly in \emph{Soviet Math. Doklady}) were 
accessible. Also I started to read papers of Chaitin~\cite{Cha66,Cha69,Cha75,Cha77} 
and others, starting with a seminal paper of Kolmogorov~\cite{Kol65} 
(in \emph{Information Transmission Problems}). Following the path of history, one 
should start with the definition of plain Kolmogorov complexity and the notion of 
randomness deficiency for a finite sequence in a finite set. 
(My opinion is that frequency definitions of randomness are now obsolete and are 
interesting only from the historical viewpoint.) All this can be called Kolmogorov's 
approach to the definition of a random finite object. Then one should give a definition 
of randomness given by Martin-L\"of for infinite sequences, and show that a difficulty 
arises when we try to characterize randomness of the sequence in terms of the plain complexity 
of its prefixes (Martin-L\"of's result saying that every sequence has prefixes where difference 
between length and complexity is large). The next step was the prefix and monotone complexities 
as invented by Levin, the theorem of Levin that characterizes randomness 
in terms of monotone or prefix complexity.\footnote{The prefix complexity and its application to 
the definition of randomness also was invented by Schnorr and Chaitin. Schnorr proposed
two different definitions of monotone complexity using monotone functions and monotone 
operators, see details in~\cite{Vyu81}.}                                                 
Finally, one should introduce the notion of a Levin 
universal semimeasure and the corresponding criterion of randomness. 

This was my plan; I had to reconstruct most of the proofs since they often were missing in original 
Levin's papers. It would be nice to have some general framework for different flavors of Kolmogorov 
complexity, but I did not succeed in doing this. While writing the paper, I got some useful feedback 
from Sasha Shen (whom I met then for the first time). Later (in his PhD thesis) he managed to provide a 
general scheme for different versions of complexity in terms of computable continuous mappings 
(that characterization was translated to a more accessible language later by Uspensky and Shen). 
 
This survey was published in the series of volumes called \emph{Semiotics and Informatics} 
in 1981~\cite{Vyu81}. Li and Vitanyi~\cite[page 193]{LiV97} wrote that ``another source for the Russian 
school is the survey by V.V.~Vyugin, \emph{Selecta Mathematica}, 
formerly \emph{Sovietica}, \textbf{13}:4~(1994), 357--389 (translated from the Russian 
\emph{Semiotika and Informatika}, \textbf{16}~(1981), 14--43).'' For Russian readers it was 
the second available survey paper after~\cite{ZvL70}, and it included main Levin's results about 
randomness with proofs.

The experienced people explained me how the Levin's papers could be cited still keeping the paper 
publisheable: references to his papers are more or less OK, 
assuming that name `Levin' never appears in the text of the paper, only in the list of references. 
(In particular, I couldn't express my gratitude to him in the text.)

When I called Kolmogorov next time (about September 1980), he said that he has read my paper and finds it 
interesting. However, he does not agree completely with the interpretation of the results. 
He said also that he would like to write a commentary and later entire paper that 
explains his views, and it would be nice to publish these papers together in one 
volume of some journal. He said also that he has read my survey and likes it 
(and later he even made a reference to it in his paper~\cite{Kol83a}).

Kolmogorov invited me to his apartment and typed (while I was there) his commentary, using his electric 
typewriter. At that time he was more and more struggling with the Parkinson disease and it was difficult 
for him to type, several letters were typed repeatedly. So I went to the office of the Logic Division of 
the Mathematics Department of Moscow State University to retype the commentary. I still keep the copy of 
the Kolmogorov's typescript. 

Kolmogorov never wrote an article he planned, and I submitted my article to the 
\emph{Information Transmission Problems journal}. The editorial office told me that they never 
accept any commentaries together with the paper. The paper was published in 1982 as~\cite{Vyu82}.

After our meeting Kolmogorov gave a talk at the seminar that was organized by him, Alexei Semenov 
and Alexander Shen, called ``Definition complexity and computational complexity'' (Slozhnost' opredelenii i 
slozhnost' vychislenii). One could say that this was a continuation of a seminar led by Kolmorogov 
and Levin many years earlier.

In his talk Kolmogorov gave the definition of an $(\alpha,\beta)$-stochastic sequence. As people remember, 
Kolmgorov had the similar notions much earlier, even before 1974 (see~\cite{CGG89}). 
I guess that my phone calls reminded him of this notion and he suggested to study its properties. 
Now this approach is developed as a part of the so-called ``algorithmic statistics'' (see~\cite{VeV2003}).

I would explain Kolmogorov's definition as follows.  Fix arbitrary non-negative numbers 
$\alpha$ and $\beta$. A string $x$ is called $(\alpha,\beta)$-stochastic if it is 
a ``$\beta$-random'' element of some ``$\alpha$-simple'' sets --- more precisely, 
if $x\in A$ and $\K(x\cnd A)\ge \log |A|-\beta$ for some finite set $A$ of strings with 
$\K(A)\le\alpha$. Here $|A|$ is the cardinality of $A$, and  $\K(x\cnd A)$ is the conditional 
Kolmogorov complexity of $x$ if $A$ is given (by list of its elements), and all logarithms are binary. 

If $x$ is not $(\alpha,\beta)$-stochastic, 
this means that there is no simple model that ``explains'' $x$ is statistical sense.

Kolmogorov asked in his talk whether non-stochastic strings exist (for reasonable values 
of $\alpha$ and $\beta$) and, if yes, how many of them can be found among strings of given length.
For such a sequence we have $d(x\cnd A)=\log |A|-\K(x\cnd A)>\beta$
for every finite set $A$ containing $x$ such that $\K(A)\le\alpha$. Let us call
$d(x\cnd A)$ the randomness deficiency of $x\in A$ with respect to a finite set $A$ see~\cite{Kol83}).
                                                                          
Later A.~Shen~\cite{She83} answered Kolmogorov's question in its simplest version: 
he gave conditions on $\alpha$ and $\beta$ that guarantee the existence of non-stochastic 
sequences and proved that the fraction of nonstochastic sequences among all strings of 
given length decreases exponentially as $\alpha$ and $\beta$ increase.\footnote{
In 1998 Levin told me that he had discussed all these issues with Kolmogorov in the early 1970s and 
and got the relevant results but not published them.}

I thought it would be important to consider not only the fraction of non-stochastic sequences 
(i.e., their measure according to the uniform distribution), but the probability to get a non-stochastic 
sequence as an output of some randomized algorithm. Formally speaking, one should find the a priori 
probability of the set of all non-stochastic strings of given length.\footnote{More correctly, we consider
the a priory measure of all continuations of strings from this set. This value is equal to 
the probability with which a probabilistic Turing machine can print an $(\alpha,\beta)$-non-stochastic
sequence (or its continuation).}

Working in this direction, I established in~\cite{Vyu85} some lower and upper bounds for the 
a priori probability of the set of all $(\alpha,\beta)$-nonstochastic strings of length $n$ 
(in terms of $\alpha$ and $\beta$).

An asymptotic version of these results can be stated as follows. Consider the a priori probability 
of the set of all $(\alpha,\beta)$-nonstochastic strings of length $n$. Here $\alpha=\alpha(n)$ 
and $\beta=\beta(n)$ are considered as functions of $n$. This measure decreases as $2^{-\alpha(n)}$ 
if $\alpha(n)$ and $\beta(n)$ are unbounded computable functions. However, if we do not require 
the functions $\alpha(n)$ and $\beta(n)$ to be computable, then (as shown in~\cite{Vyu85}) one can 
find two non-decreasing unboundeds function $\alpha(n)$ and $\beta(n)$ and a probabilistic machine 
that generates with positive (or even close to $1$) probability an infinite sequence whose $n$-bit 
prefixes are not $(\alpha(n),\beta(n))$-stochastic for infinitely many~$n$.\footnote{Here the 
function $\alpha(n)$ can be made upper semicomputable and the function $\beta(n)$ can be made computable.} 

It is easy to see that infinite sequences with this property cannot be Martin-L\"of random 
with respect to any computable measure. One can modify the argument to guarantee that machine 
generates (with positive probability) sequences that have this property and at the same time are 
not Turing-equivalent to Martin-L\"of random sequences.

One could say that the philosophical question about ``real existence'' of non-stochastic objects 
could have different answers depending on its interpretation. If we fix some computable 
bounds $\alpha(n)$ and $\beta(n)$ for non-stochasticity, the a priori probability of the set of all
non-stochastic sequences decreases exponentially. On the other hand, by choosing non-computable bounds, 
we may generate (with probability close to $1$) infinite sequences who have infinitely many non-stochastic 
prefixes.

It seems to me that Kolmogorov and Levin have different philosophical views on what 
should be the subject of algorithmic randomness theory. Kolmogorov thought that 
the most important notion of randomness deals with finite objects (binary strings), 
and should \emph{not} use the notion of measure (probability distribution). 
He believed that one should derive the properties of a string from the assumption 
that this string has maximal complexity in some class of strings~\cite{Kol83a}. 
More precisely, consider some equivalence relation on the set of $n$-bit strings. 
This relation defines equivalence classes; $n$-bit string $x$ is considered as $m$-random 
with respect to a partition if 
$\K(x\cnd D)\ge\log|D|-m$ where $D$ is the equivalence class containing $x$ (it is used as 
a condition in the conditional complexity of $x$), and $|D|$ denotes its cardinality. 
Encoding objects by their ordinal numbers, we get $\K(x\cnd D)\le\log|D|+c$ where $c$ is 
some constant. 

As an example, the definition of the $m$-Bernoulli sequence given by Kolmogorov 
in~\cite{Kol69} can serve. A finite sequence $x$ is called $m$-Bernoulli if 
$K(x|n,k)\ge\log {n\choose k}-m$, where $n$ is the length of this sequence 
and $k$ is number of ones. The element of the corresponding partition consists of all sequences 
of a given length and with a given number of ones.



This approach was developed by Eugene Asarin (one of the last Kolmogorov's students), 
its main findings are presented in his unpublished thesis (see also~\cite{Asa87}).

Levin believed that the algorithmic randomness theory should be applicable both to finite and 
infinite objects, and the notion of randomness depends on the assumed measure, so measure should 
be used. Therefore he used monotone complexity that is better suited to the definition of randomness 
for infinite sequences. For the same reason he considered infinite sequences and their degrees 
when discussing non-stochasticity.

It is possible that this philosophical difference (what is more important: finite or infinite objects) 
was meant by Kolmogorov when he said that he disagrees with the interpretation of results from my paper. 
And, indeed, he always considered finite objects when speaking about non-stochasticity.

After Kolmogorov's talk mentioned above the interest in algorithmic information theory revived 
in the USSR. Kolmogorov, Uspensky and Semenov had several good students that worked in this field, 
including Asarin, Vladimir Vovk, Shen, Andrei Muchnik. Uspensky, Semenov and Shen 
wrote a survey~\cite{USS90}; recently a monograph~\cite{Mono} appeared. But this period of 
the development of algorithmic information theory in the USSR and later Russia is beyond this paper.

The author is grateful to Alexander Shen for useful discussions that led to the clarification of many 
concepts and results.

\end{document}


\maketitle

\begin{abstract}
Эти воспоминания относятся к наиболее <<тёмному>> периоду развития алгоритмической
теории информации в СССР, когда после большого интереса к этой теме в 60-х годах и в самом 
начале 70-х годов число людей, интересующихся этой темой, резко уменьшилось. В это время
Л.\,А.\,Левин, незадолго перед своей эмиграцией, опубликовал серию статей, в которых он ввёл основные понятия
современной алгоритмической теории информации. Возрождение
интереса к колмогоровской сложности началось после знаменитого доклада А.\,Н.\,Колмогорова
на семинаре в МГУ (кафедра математической логики мехмата) в 1981~году. Примерно в это же время образовалась группа молодых 
последователей Колмогорова, которые позже значительно развили это направление.
\end{abstract}

\section{Леонид Левин} 

С Лёней Левиным я познакомился зимой 1971--72 учебного года, когда как аспирант первого года обучения 
я жил в Доме аспиранта и стажера МГУ (ДАС). Это было огромное общежитие, в котором по 4--5 человек в 
комнате вперемешку жили аспиранты первого года со всех  факультетов, а также стажеры, командировочные 
и много других посторонних личностей.

Лёня Левин закончил МГУ на год раньше меня в 1970 году. По указанию парткома и КГБ в аспирантуру 
его не взяли, несмотря на его известные научные достижения  (например, он уже тогда был автором 
обзора~\cite{ZvL70}) и поддержку Колмогорова,  и с тех пор он постоянно находился во временном 
положении. Кажется,  тогда Колмогоров устроил его на временную работу в МГУ в свою лабораторию стажёром, 
потом заместитель директора ИППИ И.\,А.\,Овсеевич принял Левина тоже на  временную работу в ИППИ 
(Институт проблем передачи информации). Иногда Лёню можно было видеть в очереди к директору 
Дома студентов, когда он пытался продлить свое официальное место в ДАСе. Впрочем, при наличии 
большого количества знакомых там можно было некоторое время жить и неофициально.

Лёня был знаком, наверное, со всеми аспирантами-математиками (и не только математиками),  живущими 
в этом общежитии, в том числе и со мной. Он был широко известен в научных  и молодёжных кругах Москвы, 
все просто называли его --- Лёня Левин.  Его интересы были чрезвычайно широки, трудно найти область 
науки и жизни, которой бы он не интересовался и не имел своего мнения. Лёня обсуждал каждый новый 
математический результат, который появлялся в то время.

Кроме этого, к нему постоянно приходили специалисты по разным областям прикладной математики из московских 
научно-исследовательских институтов, которые рассказывали про свои задачи и получали его консультации, 
а иногда Лёня решал их проблемы. 

В то время про колмогоровскую сложность мы с ним никогда не разговаривали.  Помню, 
что Лёня Левин получил тогда сильные результаты о сигнализирующих (ёмкости) 
вычислимых функций, связанные с теоремой Рабина о сжатии и с теоремой Блюма 
об ускорении, которые в то время много обсуждались.
Его статья~\cite{Lev74a} была опубликована в том же сборнике, где был помещён перевод 
на русский язык одной из статей Блюма на эту тему. Эти результаты он мне рассказывал. 

В то время (1971--1974), в аспирантуре, я занимался изучением полурешёток 
вычислимых нумераций семейств рекурсивно перечислимых множеств. Понятие вычислимой нумерации 
ввел Колмогоров, первым развивал это направление В.\,А.\,Успенский, который был моим научным 
руководителем. В 1970-х годах это направление интенсивно развивалось в Новосибирском институте 
математики, там интересовались моими работами и разбирали их на семинаре <<Алгебра и логика>>.
   
До 1975 года я не знал, что такое колмогоровская сложность и  алгоритмическая случайность. 
Правда, помню, что примерно в 1970 или 1971 году я присутствовал на многолюдном заседании 
какого-то семинара под руководством Колмогорова. По-видимому, это был прообраз
семинара по колмогоровской сложности, который он организовал в начале 1980-х годов.
В аудитории было более тридцати человек, некоторым не хватило места, и они стояли. 
Среди присутствующих, кроме Колмогорова, помню Левина, М.\,Кановича, Н.\,Петри.

Тогда уже Колмогоров обращался к Левину с большим уважением как к сравнимому с ним  по заслугам математику. 
Помню, они обсуждали какую-то задачу, которую решал Левин и кто-то сказал, что её только что решил другой 
математик. Колмогоров сказал  Левину: <<Мы с вами уже достаточно зрелые математики, чтобы не огорчаться 
из-за того,  что не мы первые решили задачу>>. 

Ещё одно воспоминание. На семинаре Колмогоров спросил Левина --- не может ли он 
сделать доклад на такую-то тему. Левин отвечал, в свойственной ему манере, 
что может сделать доклад на любую тему, в любое время, в любом месте.

Второй период знакомства с Левиным начался в 1975 году. В этом году я закончил 
аспирантуру, защитил диссертацию и был направлен (по распределению, на три года) на работу в Уфу в Башкирский
государственный университет. В Уфе были неплохие математики, но все они
занимались математическим анализом и дифференциальными уравнениями. Мне
общаться там было не с кем. 

Я часто приезжал в Москву и как-то случайно встретил Левина в МГУ. Он пригласил меня к себе на квартиру, 
которую он в то время снимал в Москве, в районе метро <<Каховская>> и Черноморского бульвара. 

В то время его положение относительно стабилизировалось, он устроился на работу в какой-то прикладной 
институт, где от него не требовали постоянного присутствия, получал зарплату, снял квартиру. 
В то же время в его жизни чувствовалось напряжение --- он понимал, что в любое время КГБ может 
в очередной раз вмешаться в его жизнь,  сделать так, чтобы его уволили с работы и лишили места жительства. 

Лёня любил гостей и был не против, когда они некоторое время жили у него.  Я думаю, что очень много 
людей близкого возраста помнят эту квартиру, его гостеприимство, веселые компании и интересные разговоры
(часто оппозиционные к власти, как это тогда было принято в среде интеллигенции). 
Часто в компании были иностранцы. Пример одного смешного разговора: говорили о том, что такое свобода и 
про несвободу в СССР.  Американка Джуди (которой очень нравилось в таких компаниях и которая несколько 
раз приезжала в СССР) заметила, что свобода --- это внутреннее ощущение, на что Лёня привел такую метафору. 
Петер Гач (его венгерский коллега; сейчас они оба работают в Boston University) приезжает в Москву с 
очень вкусной венгерской колбасой и каждый день отрезает от нее кусочек. Когда колбаса заканчивается, 
Петер возвращается в Венгрию (которая тогда воспринималась как немного <<менее социалистическая>> страна). 

Летом 1975 года я надолго приехал в Москву, и почти весь июль мы гуляли по Битцевскому лесопарку и 
разговаривали обо всём: о политике, философии, математике, про универсальные предсказания и универсальную 
полумеру, про понятие количества информации для бесконечных последовательностей. Иногда к нам присоединялся 
Альберт Абрамович Мучник.

Несмотря на то, что Левин обычно соображал значительно быстрее своего собеседника и всегда мог убедить 
его в своей правоте, разговаривать с ним было легко. Он никогда не навязывал свое мнение --- он просто 
путем рациональных рассуждений в процессе дискуссии убеждал собеседника в своей правоте, нисколько 
не унижая его. Эта особенность  привлекала к нему людей, делала их его последователями. По-видимому, 
эти  его качества понимали в парткоме и в КГБ, боялись его влияния и пытались его изолировать.  
В СССР таким людям не было места.

Тогда же (в 1975 году) Лёня сформулировал задачу, которой он придавал 
принципиальное значение. После некоторых упрощений она свелась к следующему. 
Рассматриваются все бесконечные двоичные последовательности с точностью до тьюринговой 
эквивалентности (тьюринговы  степени). Если две последовательности переводятся друг 
в друга вычислимыми операторами, можно считать, что они содержат одну и ту же 
информацию (с точностью до задания этих вычислимых операторов). Мы хотим рассматривать 
свойства (борелевские множества) бесконечных последовательностей, инвариантные относительно 
таких перекодировок (множества тьюринговых степеней или <<информационные массивы>>). Одно из таких 
множеств порождено случайными по Мартин-Лёфу последовательностями. Ранее Левин показал, что 
всякая последовательность, случайная относительно вычислимой меры, эквивалентна некоторой последовательности, 
случайной относительно равномерной меры. Это не относится к вычислимым последовательностям, 
которые тоже случайны относительно мер, сконцентрированных на них. Так что все невычислимые 
случайные по Мартин-Лёфу относительно вычислимых мер последовательности попадают в один 
инвариантный класс. Другой инвариантный класс образуют вычислимые последовательности. 
Возникает вопрос, а есть ли другие нетривиальные инвариантные классы. 
 
Конечно, этот вопрос требует уточнения. Оба рассмотренных инвариантных класса состоят 
из последовательностей, которые отражают некоторую реальность --- в том смысле, что их можно генерировать с 
помощью алгоритмов или случайных процессов. Можно предположить, что существуют искусственно 
построенные инвариантные классы, элементы которых нельзя генерировать с
 помощью алгоритмических и случайных процессов, точнее, вероятность получить такой
элемент равна нулю. Например, таким классом является тьюрингова степень одной невычислимой 
последовательности (см. известную теорему Де Леу -- Мура -- Шеннона -- Шапиро~\cite{LMSS}). 
Другие примеры таких классов были построены позже. Левин описал все такие классы. А именно, для каждого 
такого класса существует единственная последовательность, о которой
каждая последовательность из этого класса имеет бесконечное количество информации.
Кроме того, множество всех последовательностей, содержащих бесконечное количество информации
о заданной невычислимой последовательности, обладает этим свойством 
(см.~\cite{Lev74}).\footnote{Л.А. Левин предложил в этой и других более поздних статьях 
(см., например,~\cite{Lev84}) несколько определений понятия количества информации 
в одной бесконечной последовательности о другой бесконечной последовательности.
Количество информации $I(\alpha:\beta)$ должно удовлетворять свойствам инвариантности: 
эта величина не должна возрастать в случайных и детерминированных (рекурсивных) 
преобразованиях последовательностей $\alpha$ и $\beta$ (подробнее см. в~\cite{Lev84}).
Он считал, что понятие количества информации для бесконечных последовательностях
требует дальнейшего уточнения и развития.} 
Универсальная полувычислимая (априорная) мера подобного инвариантного множества равна 
нулю.\footnote{Универсальноая 
(априорная) мера была введена в работе~\cite{ZvL70}. Априорная мера множества бесконечных 
последовательностей равна вероятности того, что соответствующая вероятностная машина Тьюринга 
выдаст последовательность из этого множества.} Левин назвал такие множества \emph{пренебрежимыми}. 

Мы хотим игнорировать все свойства, образующие пренебрежимые множества. Для этого факторизуем 
все инвариантные множества по отношению эквивалентности: два таких множества эквивалентны, 
если их симметрическая разность образует пренебрежимое множество. Получим нашу основную 
структуру --- алгебру инвариантных свойств. Говоря об алгебре, мы имеем в виду теоретико-множественные операции. 
Класс эквивалентности пренебрежимых множеств образует нулевой элемент этой алгебры $\mathbf{0}$. 
Все последовательности образуют наибольший элемент $\mathbf{1}$. Другие примеры элементов: класс $\mathbf{r}$, 
порождённый случайными последовательностями, класс $\mathbf{c}$, порождённый вычислимыми последовательностями. 

Левин рассказал мне, что элементы $\textbf{c}$ и $\mathbf{r}$ являются атомами алгебры инвариантных 
свойств --- в том смысле, что не могут быть представлены в виде объединения других ненулевых элементов.

Левин назвал последовательности, случайные по Мартин-Лёфу относительно вычислимых мер, 
правильными (см.~\cite{ZvL70}, определение 4.4) или полными (см.~\cite{LeV77,Lev84}).
Последовательности, которые эквивалентны по Тьюрингу полным последовательностям, называются
регулярными (см.~\cite{Lev84}). 

Каждая регулярная последовательность либо вычислима, либо эквивалентна по Тьюрингу некоторой 
последовательности, случайной по Мартин-Лёфу относительно равномерной меры (см.~\cite{ZvL70}. 
Поскольку для каждой случайной по Мартин-Лёфу последовательности 
относительно равномерной меры последовательности
монотонная колмогоровская сложность всех её начальных фрагментов равна их длине (с точностью до
аддитивной константы), можно сказать, что информация, содержащаяся в регулярной последовательности, 
допускает оптимальное сжатие.   
Свойство элементов $\mathbf{c}$ и $\mathbf{r}$ быть атомами означает, что единственным инвариантным 
свойством регулярных последовательностей является количество информации --- конечное или бесконечное. 
Про элементы, образующие атом $\mathbf{r}$, можно сказать, что информация в наиболее плотной своей
кодировке неотличима от случайного шума.

Левин задался вопросом: исчерпывают ли эти два элемента  (порождённые множествами случайных  и вычислимых 
последовательностей) все возможные инвариантные свойства информационных массивов: верно ли, 
что $\mathbf{c}\cup \mathbf{r}=\mathbf{1}$ или есть и другие элементы?  

Летом 1975 года по возвращению в Уфу мне удалось ответить на этот вопрос, построив вычислимый оператор, 
который с близкой к 1 вероятностью выдавал невычислимые последовательности, 
не эквивалентные по Тьюрингу никакой случайной последовательности. Тем самым было доказано, 
что $\mathbf{c}\cup \mathbf{r}\ne\mathbf{1}$ и алгебра инвариантных свойств не исчерпывается 
приведёнными выше четырьмя элементами (см.~\cite{Vyu76}). 

Этот результат можно интерпретировать так, что можно с положительной вероятностью  генерировать носители 
информации, которые имеют <<нестохастический>> характер, т.е. лежат за пределами применимости теории 
вероятностей. Такие последовательности не являются регулярными и не допускают оптимального 
сжатия без потери бесконечного количества информации. Это типичный результат применения общей теории 
алгоритмов к основаниям математики (и теории вероятностей). 

Как было замечено ранее, элементы $\mathbf{c}$ и $\mathbf{r}$ являются атомами алгебры инвариантных свойств, 
в том смысле, что не могут быть представлены в виде объединения нетривиальных элементов. 
Левин задал ещё один вопрос: есть ли другие атомы в алгебре 
инвариантных свойств? Из определения легко следует, что атомов не более чем счётное число. 
Примерно в 1977--1978 годах я построил счётное семейство атомов и бесконечно делимый 
элемент (он не содержит ни одного атома). Таким образом, имеет место представление 
$$
\mathbf{c}\cup \mathbf{r}\cup\bigcup_{i=1}^\infty \mathbf{a}_i\cup \mathbf{d}=\mathbf{1},
$$
где $\mathbf{a}_i$ --- атомы, порожденные неслучайными последовательностями, 
а $\mathbf{d}$ --- остаток, который является бесконечно делимым элементом. 
Данная работа, к сожалению, не имела дальнейшего продолжения, свойства новых элементов алгебры 
инвариантных свойств остаются не изученными.

В то время Левин разрабатывал абстрактную теорию случайности индивидуальных объектов. Он рассказывал 
о попытках дать с помощью теории алгоритмов определение равномерного дефекта случайности относительно 
полумер (мер на множестве конечных и бесконечных последовательностей) так, чтобы любая бесконечная  
последовательность была бы случайной относительно априорной полумеры.\footnote{Некоторые результаты в 
этом направлении были позже опубликованы в~\cite{Lev84}.}
Он также думал о построении универсального языка для компактной записи всех математических объектов 
и функций. В таком языке константы, возникающие  в основных неравенствах для колмогоровской сложности, 
были бы небольшими. Все эти вопросы мы бесконечно обсуждали в 1975--1978 годах.

Лёня также рассказал мне про проблему перебора (P=NP), которую он сформулировал 
в заметке~\cite{Lev73}, опубликованной в журнале \emph{Проблемы  
передачи информации}, и ставшей потом знаменитой. Нужно заметить, что в отличие от Кука, Левин не только  
сформулировал проблему и привел примеры NP-полных задач, 
но и предложил алгоритм (L-search), решающий  
эту проблему за наименьшее время (только время работы этого алгоритма до сих пор не известно, 
это и есть проблема перебора). Алгоритм был основан на использовании нового вида сложности, в 
которой минимизируется не только длина программы, но и логарифм времени её работы.
Как обычно, заметка Левина была максимально краткой и не содержала доказательств --- и даже описания 
алгоритма: то, что он использует новый вид сложности, понять из неё нельзя. 
История попыток решения проблемы перебора в СССР изложена в прекрасном обзоре
Бориса Трахтенброта~\cite{Tra84}. Про алгоритм и соответствующую сложность Левин мне рассказывал.\footnote{ 
Позже этот алгоритм был представлен в работе~\cite{Lev84}, 
алгоритм Левина также приведен в обзорной статье~\cite{Vyu99}. }

Левин советовал заняться решением проблемы перебора. Тогда эта проблема не казалась такой неразрешимой, 
как сейчас. Может быть, к счастью, личная жизнь отвлекла меня от этой проблемы.

Дружеское общение с Левиным продолжалось до 1978 года, когда он решил  эмигрировать в США с помощью визы 
в Израиль. Нужно заметить, что Левин удивительно долго не собирался эмигрировать. 
Куратор из КГБ еще в начале 70-х советовал ему уехать из Москвы куда-нибудь подальше,  например, 
в Армению. Можно было понять это как намек на эмиграцию. 

В те времена отъезд за границу был безвозвратным. Отъезжающий подавал  заявление в Верховный  
Совет СССР с просьбой о лишении его советского гражданства (кроме того, за эту процедуру надо было 
довольно много заплатить). Эмигранта ожидали языковые трудности. Английский мы знали на уровне возможности 
разобрать математическую статью со словарём. Лёня тонко чувствовал русский язык, например, он был знатоком 
поэзии Александра Блока и мог часами наизусть декламировать его стихи. 

Во время нашего общения в 1975--1978 годах Лёня находился в относительной научной изоляции,  
его не приглашали делать доклады на семинарах и конференциях. Диссертацию ему не позволили защитить, 
провалив в Новосибирске, несмотря на положительные отзывы рецензентов и руководителя (Колмогоров). 
См. подробнее в~\cite{BS2009}. Сама диссертация сохранилась и сейчас опубликована 
(в английском переводе) как~\cite{Lev2010}. 

Однажды Левина пригласили на международную конференцию по компьютерным наукам в Чехословакию (Mathematical 
Foundations of Computer Science). Официально поехать на неё не было возможности ни у него, ни у меня. 
(Левина пригласил кто-то из членов оргкомитета. Он предлагал ехать мне, но  в Уфе мне сказали, чтобы 
я и не пытался --- для этого надо 18 бумаг подписать.) Статью мы туда послали без всякого разрешения 
(<<актов экспертизы>>, говоря советским языком), и она была опубликована (см.~\cite{LeV77}). За это 
могли уволить с работы и привлечь к ответственности, но советская система уже дряхлела и не была 
способна отследить все такие нарушения.
 
Конечно, у Левина было много друзей, с которыми можно было обсуждать научные вопросы:  
среди них вспоминаю Сашу Звонкина, Альберта Абрамовича Мучника, Петера Гача. И конечно, его всегда 
поддерживал Колмогоров, который представлял его статьи в \emph{Доклады Академии наук}, его статьи 
также печатали в журнале \emph{Проблемы передачи информации}.

Однако, как заметил сам Левин, это была деятельность частных лиц, которая происходила вне рамок 
какой-либо научной организации. Вопросы алгоритмической теории информации не обсуждались в университете,
в частности, Левин не мог руководить студентами. В то время алгоритмическая теория информации
была ``подозрительной'' наукой. 

Можно заметить, что наибольшую публикационную активность Лёня 
проявлял в те периоды, когда у него возникали жизненные трудности. Вот некоторые статьи опубликованные 
Левиным в эти годы;  в них представлены его основные результаты в области алгоритмической теории 
информации:~\cite{Lev73a,Lev74,Lev76,Lev76a,Lev76b,Lev76c,Lev77}. 

По вопросам, связанным с алгоритмической теорией информации, других  сотрудников в СССР, кроме меня, 
тогда у Левина не было. Из Венгрии регулярно приезжал Гач, он получил значительные результаты 
в алгоритмической теории  информации.

Я думаю, что мне повезло с тем, что у Лени Левина было так много времени для общения со мной. Если бы 
он был официально признанным в СССР знаменитым ученым (которым он с мировой точки зрения и являлся), 
то он был бы очень занят и наши встречи  были бы более короткими и формальными. 

В тоталитарном СССР наши начальники по месту работы не понимали, чем мы занимаемся,  да и не очень 
этим интересовались. Поскольку у нас не было особых карьерных возможностей  и должностей, мы были 
духовно свободными людьми (здесь я присоединяюсь к Джуди),  мы думали о чём хотели, мы не обязаны  
были регулярно печатать статьи, у нас не было дедлайнов и отчётов. Мы могли подолгу заниматься очень 
абстрактными (почти философскими) задачами. И, конечно, мы  много бездельничали.   

В 1977 году я женился и переехал в Москву. В смысле работы Москва встретила меня неприветливо. 
Преподавать в учебные заведения не брали, Москва была переполнена высококвалифицированными специалистами, 
которые ни за что не хотели уезжать в провинцию. Мне пришлось устраиваться в отраслевые институты, 
сначала в один, потом в другой, а потом и в третий. В первых двух институтах я продержался по году, 
работа там была совсем не организована, сотрудники ничего не делали, как правило, их специальность 
не соответствовала тематике института, а начальники давали идиотские задания. Впрочем, там мне как 
кандидату наук неплохо платили. В это время я, не обращая внимания на окружающую обстановку,  
дописывал статью про алгебру инвариантных свойств. 

В 1978 году Лёня и его жена Лариса получили разрешение на эмиграцию и уехали поездом в Вену, 
толпа друзей провожала их на  Белорусском вокзале. На прощание Лёня сказал мне, что после его 
отъезда алгоритмическая теория информации начнет наконец развиваться в СССР.  Он также 
написал отзыв и посоветовал мне показать статью про алгебру инвариантных свойств Колмогорову 
и дал мне его телефон.

\begin{center}
\includegraphics[width=0.9\textwidth]{levin-departure-photo.jpg}\\
Л.Левин (3-ий слева) с женой Ларисой (8-слева) и их друзьями перед эмиграцией (также
Э.Думанис (6) и его жена (4), В.Вьюгин (5) и его жена Лена (7), 
Альберт Абрамович Мучник (9))
\end{center}

\section{Андрей Николаевич Колмогоров}

Когда моя статья была готова, где-то в 1979 году, я позвонил Колмогорову и  сказал, что Лёня Левин 
посоветовал показать ему мою статью про алгоритмическую  теорию информации. Колмогоров сказал 
свой адрес и сказал, когда прийти. В назначенное время он сам открыл мне дверь, спросил, кто 
я такой --- и взял статью. Он велел мне позвонить ему через две недели. Через две недели я позвонил. 
Он стал извиняться, что не успел ещё прочитать, и попросил позвонить еще через две недели. 
Так продолжалось примерно полтора года. Я не всегда звонил каждые две недели, были перерывы на 
несколько месяцев, когда он болел. 

Примерно, в начале 1980 года интерес к колмогоровской сложности и в самом деле возник.  
Мой научный руководитель Успенский попросил меня написать обзорную статью про колмогоровскую сложность 
и алгоритмическую случайность. Я никогда ранее не систематизировал свои знания по  алгоритмической теории 
информации, но что-то осталось от прогулок с Левиным, были также его статьи 1973--1977 годов 
в \emph{Докладах Академии Наук}. Правда, там не было доказательств.  Кроме этого, я стал просматривать 
статьи Г.\,Чейтина (см.~\cite{Cha66,Cha69,Cha75,Cha77}). Начал я со статьи Колмогорова~\cite{Kol65} 
в журнале \emph{Проблемы передачи информации}. В исторической последовательности: сначала было определение 
простой колмогоровской сложности и понятия дефекта случайности конечной последовательности 
(мне кажется, что частотные определения случайности имеют только исторический интерес). 
Это назовем колмогоровским подходом к определению случайности конечного объекта. Потом надо 
было привести определение случайности по Мартин-Лефу для бесконечных последовательностей и результат 
о невозможности описать эту случайность с помощью простой колмогоровской сложности (теорему Мартин-Лёфа 
о том, что у любой последовательности есть начальные отрезки, где разница между длиной и сложностью 
сколь угодно велика). Потом шли префиксная и монотонная сложности (по Левину), описание в их терминах 
понятия случайности по Мартин-Лёфу\footnote{Префиксная сложность также была введена Шнорром и Чейтиным
и применена ими для определения случайной бесконечной последовательности. 
Шнорр предложил два различных определения монотонной сложности,
использующие монотонные функции и монотонные операторы, см.~\cite{Vyu81}.}                                
и, наконец, универсальная полумера (по Левину) и определение случайной 
последовательности в её терминах. Приходилось восстанавливать доказательства всех утверждений. 
Было бы естественно изложить все определения сложности с каких-нибудь общих позиций, но мне не 
удалось это сделать. Мне помогла критика Саши Шеня, с которым я тогда впервые познакомился. 
Многие его советы помогли улучшить статью. Позже в своей диссертации Шеню удалось представить 
различные определения колмогоровской сложности с общей топологической точки зрения.
 
Обзор был опубликован в сборнике ``Семиотика и информатика'' в 1981г.(~\cite{Vyu81}) и стал, как пишут 
Ли и Витаньи~\cite[с.~197]{LiV97}, одним из источников о советских работах по колмогоровской сложности 
(а для русскоязычных читаталей~--- вторым обзором после~\cite{ZvL70}, и в него вошли с доказательствами 
главные результаты Левина о случайности).

Авторитетные люди объяснили мне правила цитирования работ Л.А.Левина --- они были вполне 
сносными: ссылаться  можно, но нельзя, 
чтобы его фамилия встречалась в тексте статьи, в частности,  нельзя объявлять благодарность. 

Однажды, в сентябре 1980г., в ответ на мой очередной звонок, Колмогоров сказал, что он прочитал мою 
статью, всё это интересно, но он не совсем согласен с интерпретацией результатов. 
Он также сказал, что хочет составить отзыв и написать свою статью со своим взглядом 
на эту проблему. Надо напечатать, продолжал он, эти статьи вместе в одном выпуске какого-нибудь журнала.  
Он также сказал, что прочитал мой обзор и он ему понравился (позже он даже сослался 
на него в своей статье~\cite{Kol83a}).

Колмогоров пригласил меня к себе на квартиру и сам при мне напечатал отзыв на электрической пишущей 
машинке. Тогда вследствие нарастающей болезни координация  движений у него была не очень хорошая, 
некоторые буквы пропечатывались по два три раза. Мне пришлось пойти на кафедру и перепечатать 
отзыв Колмогорова (отзыв до сих пор хранится у меня). Кажется, он предварительно давал эту статью 
А.К.Звонкину, который также написал положительный отзыв.

Свою статью Колмогоров так и не написал, а мою статью я сдал в журнал \emph{Проблемы передачи информации}. 
В редакции мне сказали, что никакие отзывы вместе со статьей они не принимают. Статья была опубликована 
в 1982 году~\cite{Vyu82}.

После нашего разговора Колмогоров сделал доклад на возродившемся тогда по его инициативе семинаре 
по сложности. Он назывался <<Сложность определений и сложность вычислений>>; его руководителями 
были сам Колмогоров,  А.\,Л.\,Семенов и Шень. В каком-то смысле это было реинкарнацией его прежних 
семинаров по сложности с участием Левина.

В этом докладе Колмогоров предложил определение $(\alpha,\beta)$-стохастической последовательности. 
Судя по воспоминаниям современников, Колмогоров придумал схожие понятия гораздо раньше, ещё 
до 1974 года (см.~\cite{CGG89}). 
Получается, что смысл моих звонков заключался в том, что благодаря созданному мною беспокойству 
Колмогоров вспомнил свое определение и поставил задачу изучения этого понятия. После его доклада 
возникло направление, связанное с этим понятием, которое развивается и в настоящее время 
(алгоритмическая статистика, см.~\cite{VeV2003}).

В моем изложении определение Колмогорова выглядит следующим образом. Заданы произвольные  неотрицательные 
числа $\alpha$ и $\beta$. Конечная двоичная последовательность (строка битов) 
$x$ называется $(\alpha,\beta)$-стохастической,  если она является <<$\beta$-случайным>> 
элементом некоторого <<$\alpha$-простого>> множества $A$. Точнее, конечная двоичная 
строка $x$ называется $(\alpha,\beta)$-стохастической, 
если $x\in A$ и $\K(x\cnd A)\ge \log |A|-\beta$ для некоторого конечного множества $A$ 
сложности $\K(A)\le\alpha$; здесь $|A|$ -- число элементов множества $A$, а
$\K(x\cnd A)$ --- условная колмогоровская сложность строки $x$ относительно 
конечного множества строк $A$; здесь и далее логарифмы по основанию 2. 
Те последовательности, для которых это неверно, называются 
$(\alpha,\beta)$-нестохастическими. Для таких последовательностей невозможно подобрать 
простую стохастическую модель, объясняющую их поведение.

Колмогоров в своем докладе поставил вопрос о существовании $(\alpha,\beta)$-нестохастических 
последовательностей и о частоте их встречаемости среди всех последовательностей заданной длины.
Если $x$ -- нестохастическая последовательность, то $d(x\cnd A)=\log |A|-\K(x\cnd A)>\beta$ 
для любого конечного множества $A$, если $x\in A$ и $\K(A)\le\alpha$. 
Называем величину $d(x\cnd A)$ дефектом случайности $x\in A$ относительно конечного множества 
$A$ (см.~\cite{Kol83}). 

Позже Шень~\cite{She83} ответил на вопрос Колмогорова: он привел условия 
на $\alpha$ и $\beta$, при которых такие последовательности существуют и показал, 
что их доля экспоненциально убывает в зависимости от $\alpha$ и $\beta$.\footnote{
В 1998 г. Левин рассказал мне, что он обсуждал все эти вопросы
с Колмогоровым еще в начале 70-х годов и получил соответствующие результаты, но не публиковал их.
} 

Мне казалось важным выяснить не только частоту $(\alpha,\beta)$-нестохастических последовательностей, 
но и оценить вероятность генерации таких последовательностей с помощью вероятностных алгоритмов. 
На математическом языке это означает, что надо оценивать универсальную (априорную) полувычислимую 
меру множества всех $(\alpha,\beta)$-нестохастических последовательностей
заданной длины.\footnote{Точнее, универсальную полувычислимую меру множества
всех последовательностей, продолжающих элементы этого множества.
Эта величина равна вероятности, с которой соответствующая вероятностная машина Тьюринга
может напечатать $(\alpha,\beta)$-нестохастическую последовательность (или ее продолжение). 
}

В~\cite{Vyu85} были получены верхние и нижние оценки оценки априорной полувычислимой меры 
множества $(\alpha,\beta)$-нестохастических последовательностей длины $n$ в зависимости 
от $\alpha$ и $\beta$.

Асимптотический вариант этих результатов выглядит следующим образом. Можно оценивать в зависимости 
от $n$ априорную полувычислимую меру всех $(\alpha,\beta)$-нестохастических последовательностей 
длины $n$, где $\alpha=\alpha(n)$ и $\beta=\beta(n)$ -- некоторые функции. Тогда априорная 
полувычислимая мера множества всех таких последовательностей длины $n$ убывает 
примерно как $2^{-\alpha(n)}$ 
в том случае, если $\alpha(n)$ и $\beta(n)$ --- неограниченные вычислимые функции. Однако если отказаться 
от требования вычислимости этих функций, то, как было показано в~\cite{Vyu85}, для некоторых неограниченных 
неубывающих функций $\alpha(n)$ и $\beta(n)$ можно с близкой к единице вероятностью генерировать 
(с помощью вероятностной машины Тьюринга) бесконечные последовательности, у которых бесконечно много 
начальных фрагментов длины $n$ являются $(\alpha(n),\beta(n))$-нестохастическими 
последовательностями.\footnote{При этом функция $\beta(n)$ может быть вычислимой, 
а функция $\alpha(n)$ --- перечислимой сверху.}

Легко видеть, что такие бесконечные последовательности не могут
быть случайными по Мартин-Лёфу относительно какой-либо вычислимой
меры. Можно также добиться, чтобы они не были эквивалентны по Тьюрингу
случайным по Мартин-Лёфу последовательностям.

Таким образом, вопрос о существовании нестохастических объектов не имеет однозначного решения. 
Если задать вычислимые границы для нестохастичности, то априорная мера множества всех таких 
последовательностей экспоненциально убывает, а если границы невычислимые, 
то можно с близкой к единице вероятностью порождать бесконечные последовательности, среди начальных 
фрагментов которых бесконечно часто встречаются нестохастические последовательности. 


Здесь следует отметить различие взглядов Колмогорова и Левина на теорию  случайности, 
точнее, на то, для каких объектов интересно её развивать. Колмогоров придавал большое 
значение изучению понятия алгоритмической случайности для конечных последовательностей 
(двоичных слов).  При этом понятие меры не должно входить в это определение. 
Основная идея Колмогорова заключалась в том, чтобы выводить вероятностные 
свойства конечной последовательности из  предположения о том, что её сложность близка к максимальному 
значению в некотором содержащем её конечном множестве небольшой сложности~\cite{Kol83a}. 
Приведём интерпретацию этой идеи в следующем виде. 
Вместо меры граничные условия на случайность задаются в виде разбиения множества 
всех конечных последовательностей длины $n$ на конечные попарно непересекающиеся 
подмножества, удовлетворяющие каким-либо ограничениям, например, с ограничениями на 
сложность. 
Конечная последовательность $x$ является $m$-случайной относительно разбиения, 
если $\K(x\cnd D)\ge\log|D|-m$, где $m$ -- натуральное число, $D$ --- тот элемент разбиения, которому 
принадлежит $x$, а $|D|$ --- число его элементов.
Из теории кодирования следует, что $\K(x\cnd D)\le\log|D|+c$ 
для любого $x\in D$, где $c$ --- константа. 

В качестве примера может служить определение $m$-бернуллиевской последовательности, данное Колмогоровым 
в работе~\cite{Kol69}. Конечная последовательность $x$ называется $m$-бернуллиевской, если
$K(x|n,k)\ge\log{n\choose k}-m$, где $n$ длина этой последовательности, а $k$ -- число единиц.
Элемент соответствующего разбиения состоит из всех последовательностей заданной длины и с заданным числом
единиц.
В наиболее последовательном виде этот подход развивал ученик Колмогорова Е.\,А.\,Асарин,
его основные результаты представлены в неопубликованной диссертации (см. также~\cite{Asa87}).

Левин, напротив, считал, что развитая теория должна быть одинаково хорошо применима как к конечным, 
так и к бесконечным последовательностям, и что понятие меры может использоваться при изучении случайности. 
Поэтому он рассматривал понятие монотонной сложности как более приспособленное к определению бесконечной 
случайной последовательности. По этой же причине он рассматривал проблему существования нестохастических 
объектов для бесконечных последовательностей, а точнее, для порождаемых ими информационных массивов 
(тьюринговых степеней).

По-видимому, этим различием в подходах можно объяснить несогласие Колмогорова с интерпретацией результатов 
из моей статьи, а также его интерес к проблеме нестохастичности конечных объектов. 

После упомянутого выше доклада Колмогорова началось возрождение интереса к алгоритмической теории 
информации. У Колмогорова, Успенского и Семёнова появились очень сильные ученики 
(Е.\,Асарин, В.\,Вовк, А.\,Шень, Андрей Мучник). Позже к ним присоединился Н.\,К.\,Верещагин. Семинар по 
колмогоровской сложности в МГУ стал работать на регулярной основе и работает до сих пор. Появился 
обзор Успенского, Семенова, Шеня~\cite{USS90} и множество других публикаций, в частности, 
монография~\cite{Mono}. Но это уже другой период развития колмогоровского подхода к теории 
вероятностей и теории информации.

Автор благодарен Александру Шеню за полезные обсуждения, которые привели к уточнению многих понятий и
результатов.